# The Use of Information Theory in Evolutionary Biology

Christoph Adami[1,2,3]

[1]*Department of Microbiology and Molecular Genetics*
[2]*Department of Physics and Astronomy*
[3]*BEACON Center for the Study of Evolution in Action*
*Michigan State University, East Lansing, MI 48824*

**Abstract**

Information is a key concept in evolutionary biology. Information is stored in biological organism's genomes, and used to generate the organism as well as to maintain and control it. Information is also "that which evolves". When a population adapts to a local environment, information about this environment is fixed in a representative genome. However, when an environment changes, information can be lost. At the same time, information is processed by animal brains to survive in complex environments, and the capacity for information processing also evolves. Here I review applications of information theory to the evolution of proteins as well as to the evolution of information processing in simulated agents that adapt to perform a complex task.

## 1. Introduction

Evolutionary biology has traditionally been a science that used observation and the analysis of specimens to draw inferences about common descent, adaptation, variation, and selection[1,2]. In contrast to this discipline that requires fieldwork and meticulous attention to detail stands the mathematical theory of population genetics[3,4], which developed in parallel but somewhat removed from evolutionary biology, as it could treat exactly only very abstract cases. The mathematical theory cast Darwin's insight about inheritance, variation, and selection into formulae that could predict particular aspects of the evolutionary process, such as the probability that an allele that conferred a particular advantage would go to fixation, how long this process would take, and how it would be modified by different forms of inheritance. Missing from these two disciplines, however, was a framework that would allow us to understand the broad macro-evolutionary arcs that we can see everywhere in the biosphere and in the fossil record: the lines of descent that connect simple to complex forms of life. Granted, the existence of these unbroken lines—and the fact that they are the result of the evolutionary mechanisms at work—is not in doubt. Yet, mathematical population genetics cannot quantify them because the theory only deals with existing variation. At the same time, the uniqueness of any particular line of descent appears to preclude a generative principle, or a framework that would allow us to understand the generation of these lines from a perspective once removed from the microscopic mechanisms that shape genes one mutation at the time. In the last 23 years or so, the situation has changed dramatically because of the advent of long-term evolution experiments with replicate lines of bacteria adapting for over 50,000 generations[5,6], as well as in-silico evolution experiments covering millions of generations[7,8]. Both experimental approaches, in their own way, have provided us with key insights into the evolution of complexity on macroscopic time scales[6,8-14].

But there is a common concept that unifies the digital and the biochemical approach: information. That information is the essence of "that which evolves" has been implicit in many writings (although the word "information" does not appear in Darwin's *On the Origin of Species*). Indeed, shortly after the genesis of the theory of information at the hands of a Bell Laboratories engineer[15], this theory was thought to explain everything from the higher functions of living organisms down to metabolism, growth, and differentiation[16]. However, this optimism soon gave way to a miasma of confounding mathematical and philosophical arguments that dampened enthusiasm for the concept of information in biology for decades. To some extent, evolutionary biology was not yet ready for a quantitative treatment of "that which evolves": the year of publication of "Information in Biology"[16] coincided with the discovery of the structure of DNA, and the wealth of sequence data that catapulted evolutionary biology into the computer age was still half-a-century away.

Colloquially, information is often described as something that aids in decision-making. Interestingly, this is very close to the mathematical meaning of "information", which is concerned with quantifying the ability to make predictions about uncertain systems. Life—among many other aspects—has the peculiar property of displaying behavior or characters that are appropriate, given the environment. We recognize this of course as the consequence of adaptation, but the outcome is that the adapted organism's decisions are "in tune" with its environment: the organism has *information* about its environment. One of the insights that has emerged from the theory of computation is that information must be physical: information cannot exist without a physical substrate that encodes it[17]. In computers, information is encoded in zeros and ones, which themselves are represented by different voltages on semiconductors. The information we retain in our brains also has a physical substrate, even though its physiological basis depends on the type of memory and is far from certain. Context-appropriate decisions require information, however it is stored. For cells, we now know that this information is stored in a cell's inherited genetic material, and is precisely the kind that Shannon described in his 1948 articles. If inherited genetic material represents information, then how did the information-carrying molecules acquire it? Is the amount of information stored in genes increasing throughout evolution, and if so, why? How much information does an organism store? How much in a single gene? If we can replace a discussion of the evolution of complexity along the various lines of descent by a discussion of the evolution of information, perhaps then we can find those general principles that have eluded us so far.

In this review, I focus on two uses of information theory in evolutionary biology: First, the quantification of the information content of genes and proteins and how this information may have evolved along the branches of the tree of life. Second, the evolution of information-processing structures (such as brains) that control animals, and how the functional complexity of these brains (and how they evolve) could be quantified using information theory. The latter approach reinforces a concept that has appeared in neuroscience repeatedly: the value of information for an adapted organism is fitness[18], and the complexity of an organism's brain must be reflected in how it manages to process, integrate, and make use of information for its own advantage[19].

## 2. Entropy and Information in Molecular Sequences

To define entropy and information, we first must define the concept of a *random variable*. In probability theory, a random variable $X$ is a mathematical object that can take on a finite number of different *states* $x_1 \cdots x_n$ with specified probabilities $p_1, \cdots, p_n$. We should keep in mind that a mathematical random variable is a description—sometimes accurate, sometimes not—of a physical object. For example, the random variable that we would use to describe a fair coin has two states: $x_1 =$ heads and $x_2 =$ tails, with probabilities $p_1 = p_2 = 0.5$. Of course, an actual coin is a far more complex device: it may deviate from being true, it may land on an edge once in a while, and its faces can make different angles with true North. Yet, when coins are used for demonstrations in probability theory or statistics, they are most succinctly described with two states and two equal probabilities. Nucleic acids can be described probabilistically in a similar manner. We can define a nucleic acid random variable $X$ as having four states $x_1 =$ A, $x_2 =$ C, $x_3 =$ G and $x_4 =$ T which it can take on with probabilities $p_1, \cdots, p_4$, while being perfectly aware that the nucleic acid molecules themselves are far more complex, and deserve a richer description than the four-state abstraction. But given the role that these molecules play as information carriers of the genetic material, this abstraction will serve us very well going forward.

**2.1 Entropy**

Using the concept of a random variable $X$, we can define its *entropy* (sometimes called *uncertainty*) as[20,21]

$$H(X) = -\sum_{i=1}^{n} p_i \log p_i . \quad (1)$$

Here, the logarithm is taken to an arbitrary base that will normalize (and give units to) the entropy. If we choose the dual logarithm, the units are "bits", while if we choose base $e$, the units are "nats". Here, I will often choose the size of the alphabet as the base of the logarithm, and call the unit the "mer"[22]. So, if we describe nucleic acid sequences (alphabet size 4), a single nucleotide can have up to one "mer" of entropy, while if we describe proteins (logarithms taken to the base 20), a single residue can have up to one mer of entropy. Naturally, a five-mer has up to 5 mers of entropy, and so on.

A true coin, we can immediately convince ourselves, has an entropy of one bit. A single random nucleotide, by the same reasoning, has an entropy of one mer (or two bits) because

$$H(X) = -\sum_{i=1}^{4} 1/4 \log 1/4 = 1. \quad (2)$$

What is the entropy of a non-random nucleotide? To determine this, we have to find the probabilities $p_i$ with which that nucleotide is found at a particular position within a gene. Say we are interested in nucleotide 28 of the 76 base pair tRNA gene of the bacterium *Escherichia coli*. What is its entropy? To determine this, we need to obtain an estimate of the probability that any of the four nucleotides are found at that particular position. This kind of information can be gained from sequence repositories. For example, the database tRNAdb[23] contains sequences for more than 12,000 tRNA genes. For the *E. coli* tRNA gene, among 33 verified sequences (for different anticodons), we find five that show 'A' at the 28th position, 17 have a 'C', 5 have a 'G', and 6 have a 'T'. We can use these numbers to estimate the substitution probabilities at this position as

$$p_{28}(A) = 5/33, \, p_{28}(C) = 17/33, \, p_{28}(G) = 5/33, \, p_{28}(T) = 6/33, \quad (3)$$

which, even though the statistics are not good, allow us to infer that 'C' is preferred at that position. The entropy of position variable $X_{28}$ can now be estimated as

$$H(X_{28}) = -2 \times \tfrac{5}{33} \log_2 \tfrac{5}{33} - \tfrac{17}{33} \log_2 \tfrac{17}{33} - \tfrac{6}{33} \log_2 \tfrac{6}{33} \approx 1.765 \text{ bits}, \quad (4)$$

or less than the maximal 2 bits we would expect if all nucleotides appeared with equal probability. Such an uneven distribution of states immediately suggests a "betting" strategy that would allow us to make predictions with accuracy better than chance about the state of position variable $X_{28}$: If we bet that we would see a 'C' there, then we would be right over half the time on average, as opposed to a quarter of the time for a variable that is evenly distributed across the four states. In other words, information is stored in this variable.

**2.2 Information**

To learn how to quantify the amount of information stored, let us go through the same exercise for a different position (say, position 41[a]) of that molecule, to find approximately

$$p_{41}(A) = 0.24, \, p_{41}(C) = 0.46, \, p_{41}(G) = 0.21, \, p_{41}(T) = 0.09, \quad (5)$$

so that $H(X_{41}) \approx 1.795$ bits. In order to determine how likely it is to find any particular nucleotide at position 41 *given* position 28 is a 'C', for example, we have to collect *conditional* probabilities. They are easily obtained if we know the joint probability to observe any of the 16 combinations AA…TT at the two positions. The conditional probability to observe state *j* at position 41 given state *i* at position 28 is

$$p_{i|j} = \frac{p_{ij}}{p_j}, \qquad (6)$$

where $p_{ij}$ is the *joint* probability to observe state $i$ at position 28 and at the same time state $j$ at position 41 and the notation "$i|j$" is read as "$i$ given $j$". Collecting these probabilities from the sequence data gives the probability matrix that relates the random variable $X_{28}$ to the variable $X_{41}$

$$p(X_{41}|X_{28}) = \begin{pmatrix} p(A|A) & p(A|C) & p(A|G) & p(A|T) \\ p(C|A) & p(C|C) & p(C|G) & p(C|T) \\ p(G|A) & p(G|C) & p(G|G) & p(G|T) \\ p(T|A) & p(T|C) & p(T|G) & p(T|T) \end{pmatrix} = \begin{pmatrix} 0.2 & 0.235 & 0 & 0.5 \\ 0 & 0.706 & 0.2 & 0.333 \\ 0.8 & 0 & 0.4 & 0.167 \\ 0 & 0.059 & 0.4 & 0 \end{pmatrix}.$$
(7)

---

[a] The precise numbering of nucleotide positions differs between databases

We can glean important information from these probabilities. It is clear, for example, that positions 28 and 41 are not independent from each other. If nucleotide 28 is an 'A', then position 41 can only be an 'A' or a 'G', but mostly (4/5 times) you expect a 'G'. But consider the dependence between nucleotides 42 and 28:

$$p(X_{42}|X_{28}) = \begin{pmatrix} 0 & 0 & 0 & 1 \\ 0 & 0 & 1 & 0 \\ 0 & 1 & 0 & 0 \\ 1 & 0 & 0 & 0 \end{pmatrix}. \qquad (8)$$

This dependence is striking: if you know position 28, you can predict (based on the sequence data given) position 42 with certainty. The reason for this perfect correlation lies in the functional interaction between the sites: 28 and 42 are paired in a stem of the tRNA molecule in a Watson-Crick pair: to enable the pairing, a 'G' must be associated with a 'C', and a 'T' must be associated with an 'A'. It does not matter which is at any position as long as the paired nucleotide is complementary. And it is also clear that these associations are maintained by the selective pressures of Darwinian evolution: a substitution that breaks the pattern leads to a molecule that does not fold into the correct shape to efficiently translate messenger RNA into proteins. As a consequence, the organism bearing such a mutation will be eliminated from the gene pool. This simple example shows clearly the relationship between information theory and evolutionary biology: fitness is reflected in information, and when selective pressures maximize fitness, information must be maximized concurrently. We can now proceed and calculate the information content.

Each column in Eq. (7) represents a conditional probability to find a particular nucleotide at position 41 given a particular value is found at position 28. We can use these values to calculate the conditional entropy to find a particular nucleotide, given that position 28 is 'A', for example, as

$$H(X_{41}|X_{28} = A) = -0.2\log_2 0.2 - 0.8\log_2 0.8 \approx 0.72 \text{ bits}. \qquad (9)$$

This allows us to calculate the amount of information that is revealed (about $X_{41}$) by knowing the state of $X_{28}$. If we do not know the state of $X_{28}$, our uncertainty about $X_{41}$ is 1.795 bits, as calculated above. But revealing that $X_{28}$ actually is an 'A' has reduced our uncertainty to 0.72 bits, as we saw in Eq. (9). The information we obtained is then just the difference:

$$I(X_{41} : X_{28} = A) = H(X_{41}) - H(X_{41} | X_{28} = A) \approx 1.075 \text{ bits}, \quad (10)$$

that is, just over one bit. The notation in Eq. (10), indicating information between two variables by a colon (sometimes a semi-colon) is conventional. We can also calculate the *average* amount of information about $X_{41}$ that is gained by revealing the state of $X_{28}$ as

$$I(X_{41} : X_{28}) = H(X_{41}) - H(X_{41} | X_{28}) \approx 0.64 \text{ bits}. \quad (11)$$

Here, $H(X_{41} | X_{28})$ is the average conditional entropy of $X_{41}$ given $X_{28}$, obtained by averaging the four conditional entropies (for the four possible states of $X_{28}$) using the probabilities with which $X_{28}$ occurs in any of its four states, given by Eq. (3). If we apply the same calculation to the pair of positions $X_{42}$ and $X_{28}$, we should find that knowing $X_{28}$ reduces our uncertainty about $X_{42}$ to zero: indeed, $X_{28}$ carries perfect information about $X_{42}$. The covariance between residues in an RNA secondary structure captured by the mutual entropy can be used to predict secondary structure from sequence alignments alone[24].

**2.3 Information Content of Proteins**

We have seen that different positions within a biomolecule can carry information about other positions, but how much information do they store about the *environment* within which they evolve? This question can be answered using the same information-theoretic formalism introduced above. Information is defined as a reduction in our uncertainty (caused by our ability to make predictions with an accuracy better than chance) when armed with information. Here we will use proteins as our biomolecules, which means our random variables can take on 20 states, and our protein variable will be given by the joint variable

$$X = X_1 X_2 \cdots X_L, \quad (12)$$

where $L$ is the number of residues in the protein. We now ask: "How much information *about the environment* (rather than about another residue) is stored in a particular residue?" To answer this, we have to first calculate the uncertainty about any particular residue in the absence of information about the environment. Clearly, it is the environment within which a protein finds itself that constrains the particular amino acids that a position variable can take on. If I do not specify this environment, there is nothing that constrains any particular residue *i*, and as a consequence the entropy is maximal:

$$H(X_i) = H_{max} = \log_2 20 \approx 4.32 \text{ bits}. \quad (13)$$

In any functional protein, the residue is highly constrained, however. Let us imagine that the possible states of the environment can be described by a random variable $E$ (that takes on specific environmental states $e_j$ with given probabilities). Then the information about environment $E = e_j$ contained in position variable $X_i$ of protein $X$ is given by

$$I(X_i : E = e_j) = H_{max} - H(X_i | E = e_j) \quad , \qquad (14)$$

in perfect analogy to Eq. (10). How do we calculate the information content of the entire protein, armed only with the information content of residues? If residues do not interact (that is, the state of a residue at one position does not reveal any information about the state of a residue at another position), then the information content of the protein would just be a sum of the information content of each residue:

$$I(X : E = e_j) = \sum_{i=1}^{L} I(X_i : E = e_j) \quad . \qquad (15)$$

This independence of positions certainly could not be assumed in RNA molecules that rely on Watson-Crick binding to establish their secondary structure. In proteins, correlation between residues are much weaker (but certainly still important, see, e.g.[25-33].), and we can take Eq. (15) as a first-order approximation of the information content, while keeping in mind that residue-residue correlations encode important information about the stability of the protein and its functional affinity to other molecules. Note, however, that a population with two or more subdivisions, where each subpopulation has different amino acid frequencies, can mimic residue correlations on the level of the whole population when there are none on the level of the subpopulations[34].

For most cases that we will have to deal with, a protein is only functional in a very defined cellular environment, and as a consequence the conditional entropy of a residue is fixed by the substitution probabilities that we can observe. Let us take as an example the rodent homeodomain protein[35], defined by 57 residues. The environment for this protein is of course the rodent, and we might surmise that the information content of the homeodomain protein in rodents is different from the homeodomain protein in primates, for example, simply because primates and rodents have diverged about 100 million years ago[36], and have since taken independent evolutionary paths. We can test this hypothesis by calculating the information content of rodent proteins and compare it to the primate version, using substitution probabilities inferred from sequence data that can be found, for example, in the Pfam database[37]. Let us first look at the entropy *per residue*, along the chain length of the 57-mer. But instead of calculating the entropy in bits (by taking the base-2 logarithm), we will calculate the entropy in "mers", by taking the logarithm to base 20. This way, a single residue can have at most 1 mer of entropy, and the 57-mer has at most 57 mers of entropy. The entropic profile (entropy per site as a function of site) of the rodent homeodomain protein depicted in Figure 1 shows that the entropy varies considerably from site to site, with strongly conserved as well as highly variable residues.

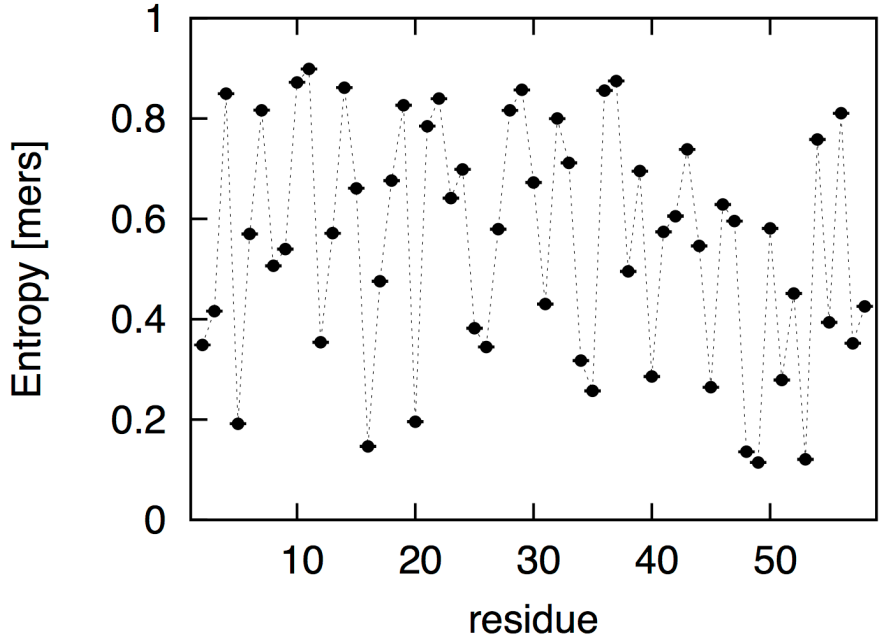

**FIG 1**: Entropic profile of the 57 amino acid rodent homeodomain, obtained from 810 sequences in Pfam (accessed February 3rd, 2011). Error of the mean is smaller than the data points shown. Residues are numbered 2-58 as is common for this domain[35].

When estimating entropies from finite ensembles (small number of sequences), care must be taken to correct for the bias that is inherent in estimating the probabilities from the frequencies. Rare sequences will be assigned zero probabilities in small ensembles but not in larger ones. Because this error is not symmetric (probabilities will always be underestimated), the bias is always towards smaller entropies. Several methods can be applied to correct for this, and I have used here the second order bias correction, described for example in Ref.[38]. Summing up the entropies per site shown in Fig. 1, we can get an estimate for the information content by applying Eq. (15). The maximal entropy $H_{max}$, when measured in mers, is of course 57, so the information content is just

$$I_{Rodentia} = 57 - \sum_{i=1}^{57} H(X_i | e_{Rodentia}) , \quad (16)$$

which comes out to

$$I_{Rodentia} = 25.29 \pm 0.09 \text{ mers} , \quad (17)$$

where the error is obtained from the theoretical estimate of the variance given the frequency estimate[38].

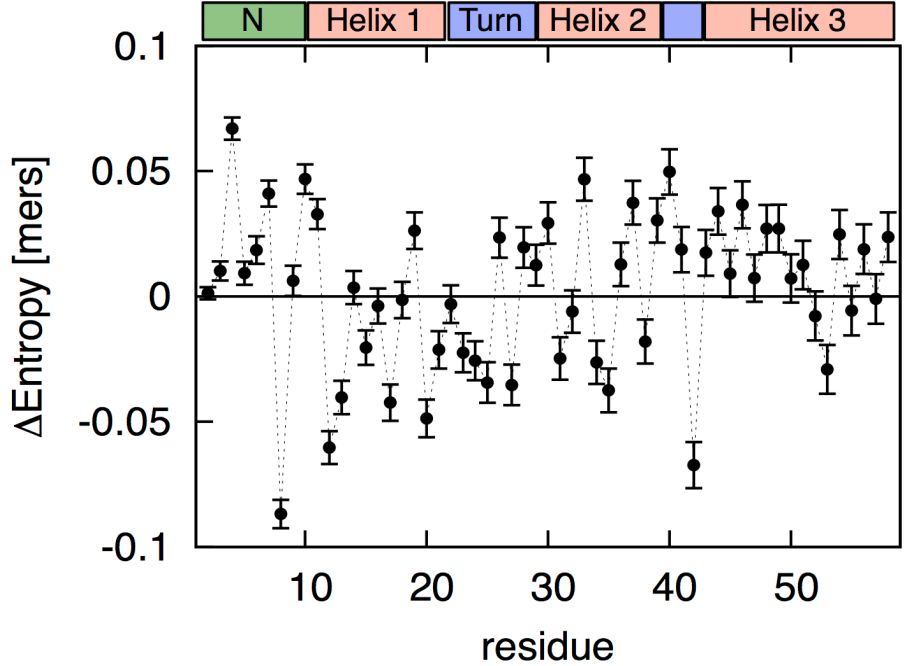

**FIG. 2**: Difference between entropic profile ΔEntropy of the homeobox protein of rodents and primates (the latter from 903 sequences in Pfam, accessed February 3rd, 2011). Error bars are the error of the mean of the difference, using the average of the number of sequences. The colored boxes indicate structural domains as determined for the fly version of this gene. ('N' refers to the protein's "N-terminus".)

The same analysis can be repeated for the primate homeodomain protein. In Fig. 2, we can see the difference between the "entropic profile" of rodents and primates

$$\Delta \text{Entropy} = H(X_i | e_{\text{Rodentia}}) - H(X_i | e_{\text{Primates}}) \ , \qquad (18)$$

which shows some significant differences, in particular, it seems, at the edges between structural motifs in the protein.

When summing up the entropies to arrive at the total information content of the primate homeodomain protein we obtain

$$I_{\text{Primates}} = 25.43 \pm 0.08 \text{ mers} \ , \qquad (19)$$

which is identical to the information content of rodent homeodomains within statistical error. We can thus conclude that while the information is encoded somewhat differently between the rodent and the primate version of this protein, the total information content is the same.

## 3. Evolution of Information

While the total information content of the homeodomain protein has not changed between rodents and primates, what about longer time intervals? If we take a protein that is ubiquitous among different forms of life (that is, its homologue is present in many different branches) has its information content changed as it is used in more and more complex forms? One line of argument tells us that if the function of the protein is the same throughout evolutionary history, then its information content should be the same in each variant. We saw a hint of that when comparing the information content of the homeodomain protein between rodents and primates. But we can also argue instead that because information is measured relative to the environment the protein (and thus, the organism) finds itself in, then organisms that live in very different environments can potentially have different information content even if the sequences encoding the proteins are homologous. Thus, we could expect differences in protein information content in organisms that are different enough that the protein is used in different ways. But it is certainly not clear whether we should observe a trend of increasing or decreasing information along the line of descent. In order to get a first glimpse at what these differences could be like, I will take a look here at the evolution of information in two proteins that are important in the function of most animals: the homeodomain protein and the COX2 (cytochrome-c-oxidase subunit 2) protein.

The homeodomain (or homebox) protein is essential in determining the pattern of development in animals: it is crucial in directing the arrangement of cells according to a particular body plan[39]. In other words, the homeobox determines where the head goes and where the tail. Although it is often said that these proteins are specific to animals, some plants have homeodomain proteins that are homologous to those I study here[40]. The COX2 protein, on the other hand, is a subunit of a large protein complex with 13 subunits[41]. While a non-functioning (or severely impaired) homeobox protein certainly leads to aborted development, an impaired COX complex has a much less drastic effect: it leads to mitochondrial myopathy due to a cytochrome oxidase deficiency[42], but is usually not fatal[43]. Thus, by testing the changes within these two proteins, we are examining proteins with very different selective pressures acting on them.

To calculate the information content of each of these proteins along the evolutionary line of descent, in principle we need access to the sequences of extinct forms of life. Even though the resurrection of such extinct sequences is possible in principle[44] using an approach dubbed "paleogenetics"[45,46], we can take a shortcut by grouping sequences according to the depth that they occupy within the phylogenetic tree. So, when we measure the information content of the homeobox protein on the taxonomic level of the family, we include in there the sequences of homeobox proteins of chimpanzees, gorillas and orangutans along with humans. As the chimpanzee version, for example, is essentially identical with the human version, we do not expect to see any change in information content when moving from the species level to the genus level. But, we can expect that by grouping the sequences on the family level (rather than the genus or species level), we move closer towards evolutionarily more ancient proteins, in particular because this group is used to reconstruct the sequence of the ancestor of that group. The great apes are but one family of the order of "Primates", which besides the apes also contains the families of monkeys, lemurs, lorises, tarsiers, and galagos. Looking at the homebox protein of all the primates then takes us further back in time. A simplified version of the phylogeny of animals is shown in Fig. 3, which shows the hierarchical organization of the tree.

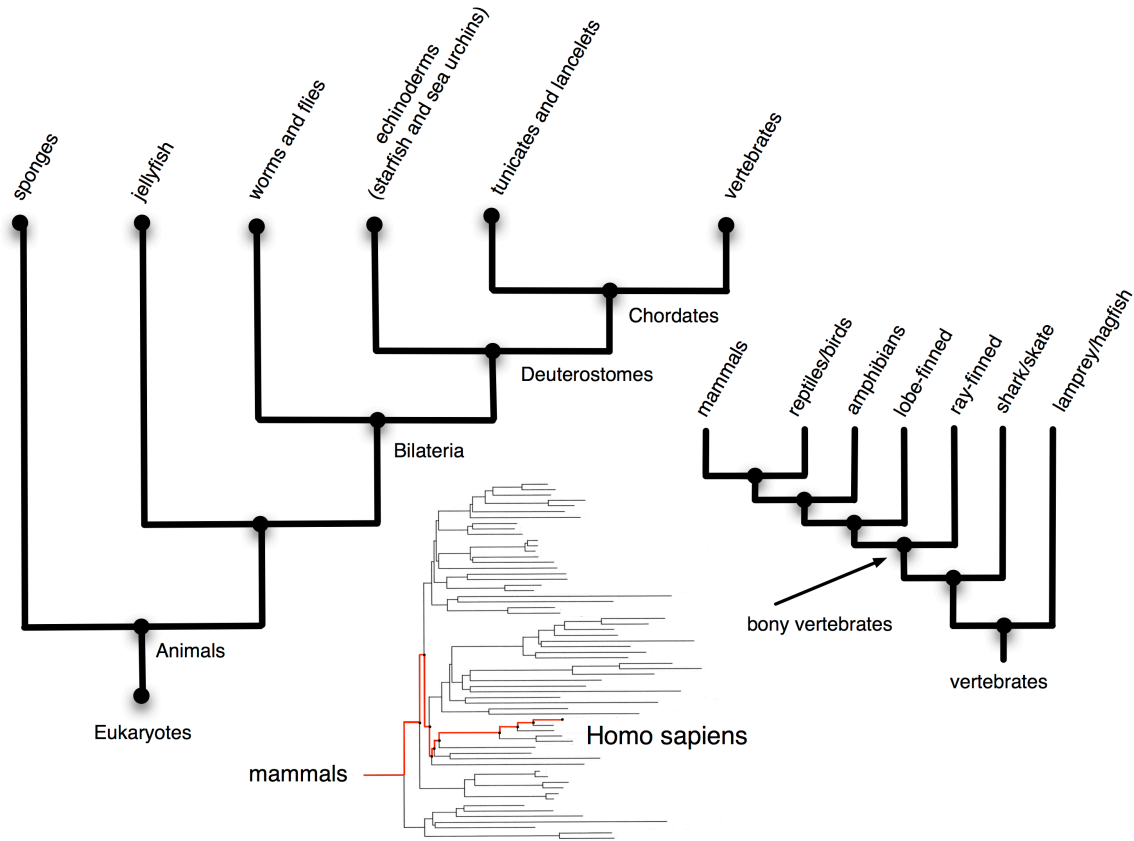

**FIG. 3**: Simplified phylogenetic classification of animals. At the root of this tree (on the left tree) are the eukaryotes, but only the animal branch is shown here. If we follow the line of descent of humans, we move on the branch towards the vertebrates. The vertebrate clade itself is shown in the tree on the right, and the line of descent through this tree follows the branches that end in the mammals. The mammal tree, finally, is shown at the bottom, with the line ending in Homo sapiens indicated in red.

The database Pfam uses a range of different taxonomic levels (anywhere from 12 to 22, depending on the branch) defined by the NCBI Taxonomy Project[47], which we can take as a convenient proxy for taxonomic depth: ranging from the most basal taxonomic identifications (such as phylum) to the most specific ones. In Fig. 4, we can see the total sequence entropy

$$H_k(X) = \sum_{i=1}^{57} H(X_i | e_k), \qquad (20)$$

for sequences with the NCBI taxonomic level $k$, as a function of the level depth. Note that sequences at level $k$ always include all the sequences at level $k$-1. Thus, $H_1(X)$, which is the entropy of all homeodomain sequences at level $k = 1$, includes the sequences of all eukaryotes. Of course, the taxonomic level description is not a perfect proxy for time. On the vertebrate line for example, the genus *Homo* occupies level $k=14$, whereas the genus *Mus* occupies level $k=16$. If we now plot $H_k(X)$ versus $k$ (for the major phylogenetic groups only), we see a curious splitting of the lines based only on total sequence entropy, and thus information (as information is just $I=57-H$ if we measure entropy in mers). At the base of the tree, the metazoan sequences split into chordate proteins with a lower information content (higher entropy) and arthropod sequences with higher information content, possibly reflecting the different uses of the homeobox in these two groups. The chordate group itself splits into mammalian proteins and the fish homeodomain. There is even a notable split in information content into two major groups within the fishes.

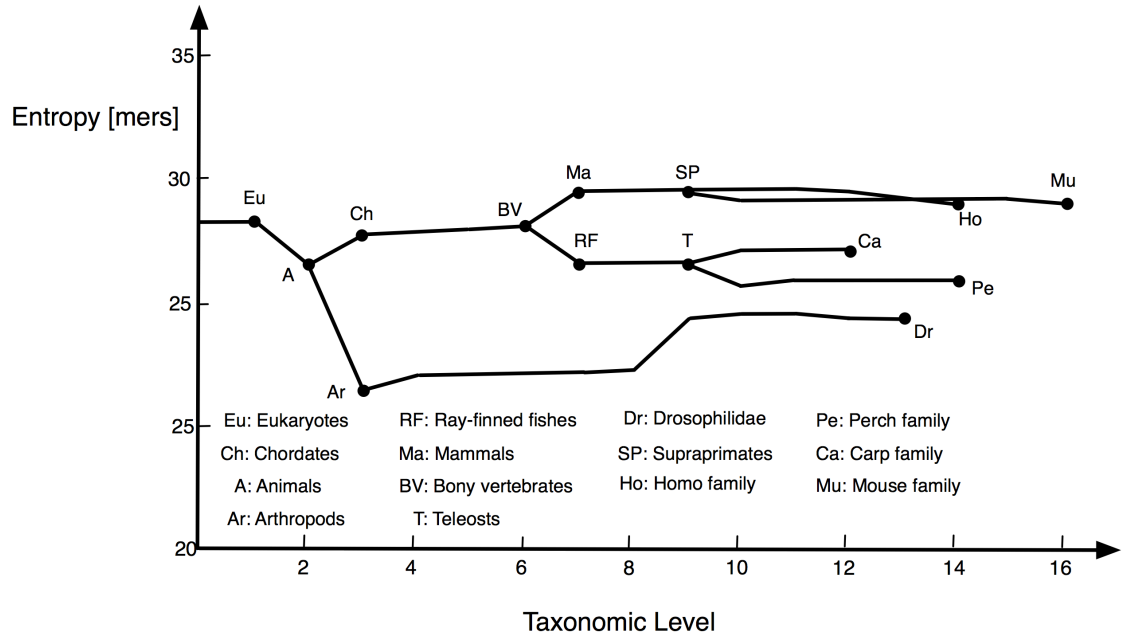

**FIG. 4**: Entropy of homeobox domain protein sequences (PF00046 in the Pfam database, accessed July 20[th], 2006) as a function of taxonomic depth for different major groups that have at last 200 sequences in the database, connected by phylogenetic relationships. Selected groups are annotated by name. 57 core residues were used to calculate the molecular entropy. Core residues have at least 70% sequence in the database.

The same analysis applied to subunit II of the COX protein (counting only 120 residue sites that have sufficient statistics in the database) gives a very different picture. Except for an obvious split of the bacterial version of the protein and the eukaryotic one, the total entropy markedly decreases across the lines as the taxonomic depth increases. Furthermore, the arthropod COX2 is more entropic than the vertebrate one (see Fig. 5) as opposed to the ordering for the homeobox protein. This finding suggests that the evolution of the protein information content is specific to each protein, and most likely reflects the adaptive value of the protein for each family.

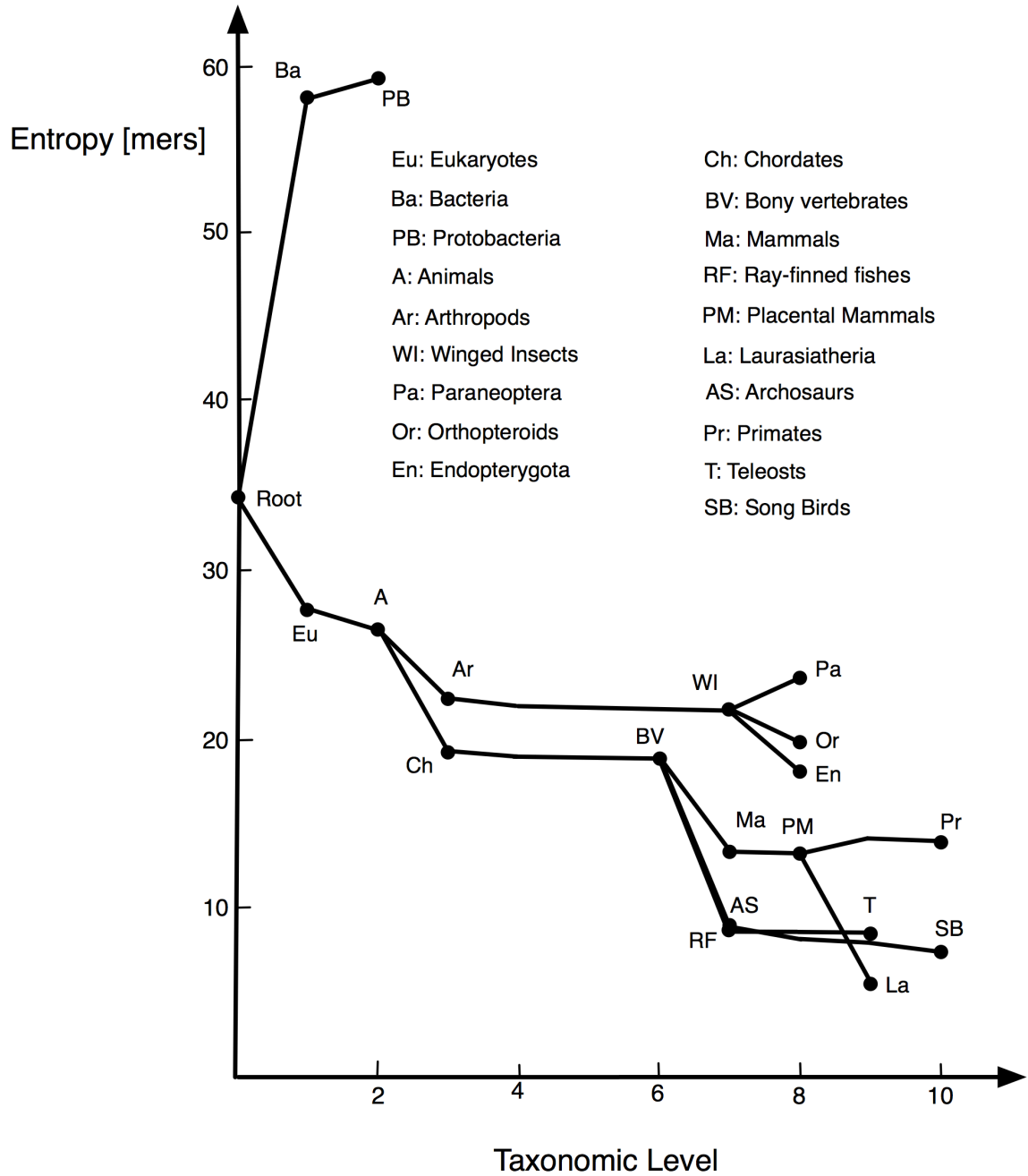

FIG. 5: Entropy of COX subunit II (PF00116 in the Pfam database, accessed June 22[nd], 2006) protein sequences as a function of taxonomic depth for selected different groups (at least 200 sequences per group), connected by phylogenetic relationships. 120 core residues were used to calculate the molecular entropy.

## 4. Evolution of information in robots and animats

The evolution of information within the genes of adapting organisms is but one use of information theory in evolutionary biology. Just as anticipated in the heydays of the "Cybernetics" movement[48], information theory has indeed something to say about the evolution of information-processing in animal brains. The general idea behind the connection between information and function is simple: because information (about a particular system) is what allows the bearer to make predictions (about that particular system) with accuracy better than chance, information is valuable as long as prediction is valuable. In an uncertain world, making accurate predictions is tantamount to survival. In other words, we expect that information, acquired from the environment and processed, has survival value and therefore is selected for in evolution.

### 4.1. Predictive information

The connection between information and fitness can be made much more precise. A key relation between information and its value for agents that survive in an uncertain world as a consequence of their actions in it was provided by Ay et al.[49], who applied a measure called "predictive information" (defined earlier by Bialek et al.[50] in the context of dynamical systems theory) to characterize the behavioral complexity of an autonomous robot. These authors showed that the mutual entropy between a changing world (as represented by changing states in an organism's sensors) and the actions of motors that drive the agent's behavior (thus changing the future perceived states) is equivalent to Bialek's predictive information as long as the agent's decisions are Markovian, that is, only depend on the state of the agent and the environment at the preceding time. This predictive information is defined as the shared entropy between motor variables $Y_t$ and the sensor variables at the subsequent time point $X_{t+1}$

$$I_{\text{pred}} = I(Y_t : X_{t+1}) = H(X_{t+1}) - H(X_{t+1} | Y_t) \ . \quad (21)$$

Here, $H(X_{t+1})$ is the entropy of the sensor states at time $t+1$, defined as

$$H(X_{t+1}) = -\sum_{x_{t+1}} p(x_{t+1}) \log p(x_{t+1}) \ , \quad (22)$$

using the probability distribution $p(x_{t+1})$ over the sensor states $x_{t+1}$ at time $t+1$. The conditional entropy $H(X_{t+1} | Y_t)$ characterizes how much is left uncertain about the future sensor states $X_{t+1}$ given the robot's actions in the present, that is, the state of the motors at time $t$, and can be calculated in the standard manner[20,21] from the joint probability distribution of present motor states and future sensor states $p(x_{t+1} | y_t)$.

As Eq. (21) implies, the predictive information measures how much of the entropy of sensorial states—that is, the uncertainty about what the detectors will record next—is explained by the motor states at the preceding time point. For example, if the motor states at time $t$ perfectly predict what will appear in the sensors at time $t+1$, then the predictive information is maximal. Another version of the predictive information studies not the effect the motors have on future sensor states, but the effect the sensors have on future motor states instead, for example in order to guide an autonomous robot through a maze[51]. In the former case, the predictive information quantifies how actions change the perceived world, while in the latter case the predictive information characterizes how the perceived world changes the robot's actions. Both formulations, however, are equivalent when taking into account how world and robot states are being updated[51]. While it is clear that measures such as predictive information should increase as an agent or robot learns to behave appropriately in a complex world, it is not at all clear whether information could be used as an objective function that, if maximized, will lead to appropriate behavior of the robot. This is the basic hypothesis of Linsker's "Infomax" principle[52], which posits that neural control structures evolve to maximize "information preservation" subject to constraints. This hypothesis implies that the infomax principle could play the role of a guiding force in the organization of perceptual systems. This is precisely what has been observed in experiments with autonomous robots evolved to perform a variety of tasks. For example, in one task visual and tactile information had to be integrated in order to grab an object[53], while in another groups of five robots were evolved to move in a coordinated fashion[54] or else to navigate according to a map[55]. Such experiments suggest that there may be a deeper connection between information and fitness that goes beyond the regularities induced by a perception-action loop, that connects fitness (in the evolutionary sense as the growth rate of a population) directly to information.

Indeed, Rivoire and Leibler[18] recently studied abstract models of the population dynamics of evolving "finite-state agents" that optimize their response to a changing environment and found just such a relationship. In such a description, agents respond to a changing environment with a probability distribution $\pi(\sigma_t | \sigma_{t-1})$ of changing from state $\sigma_{t-1}$ to state $\sigma_t$, in order to maximize the growth-rate of the population. Under fairly general assumptions, the growth rate is maximized if the Shannon information that the agents can extract from the changing environment is maximal[18]. For our purposes, this Shannon information is nothing but the predictive information discussed above (see Supplementary Text S1 in Ref.[51] for a discussion of that point). However, such a simple relationship only holds if each agent perceives the environment in the same manner, and if information is acquired *only* from the environment. If information is inherited or retrieved from memory, on the other hand, predictive information cannot maximize fitness. This is easily seen if we consider an agent that makes decisions based on a combination of sensory input and memory. If memory states (instead of sensor states) best predict an agent's actions, the correlation between sensors and motors may be lost even though the actions are appropriate. A typical case would be navigation under conditions when the sensors do not provide accurate information about the environment, but the agent has nevertheless learned the required actions "by heart". In such a scenario, the predictive information would be low because the actions do not correlate with the sensors. Yet, the fitness is high because the actions were controlled by memory, not by the sensors. Rivoire and Leibler show further that if the actions of an agent are always optimal given the environment then a different measure maximizes fitness, namely the

shared entropy between sensors and variables *given* the previous time step's sensor states[b]:

$$I_{\text{causal}} = I(X_t : Y_{t+1} | X_{t-1}) \ . \quad (23)$$

In most realistic situations, however, optimal navigation strategies cannot be assumed. Indeed, as optimal strategies are (in a sense) the goal of evolutionary adaptation, such a measure could conceivably only apply at the endpoint of evolution. Thus, no general expression can be derived that ties these informational quantities directly to fitness.

**4.2. Integrated Information**

What are the aspects of information processing that distinguish complex brains from simple ones? Clearly, processing large amounts of information is important, but a large capacity is not necessarily a sign of high complexity. It has been argued that a hallmark of complex brain function is its ability to integrate disparate streams of information and mold them into a complex *gestalt* that represents more than the sum of its parts[56-65]. These streams of information come from different sensorial modalities such as vision, sound, and olfaction, but also (and importantly) memory, and create a conscious experience in our brains that allows us to function at levels not available to purely reactive brains. One way to measure how much information a network processes is to calculate the shared entropy between the nodes at time $t$ and time $t+1$

$$I_{\text{total}} = I(Z_t : Z_{t+1}) \ . \quad (24)$$

Here, $Z_t$ represents the state of the entire network (not just the sensors or motors) at time $t$, and thus the total information captures information processing among all nodes of the network, and can in principle be larger or smaller than the predictive information that only considers processing between sensor and motors.

We can write the network random variable $Z_t$ as a product of the random variables that describe each node *i*, that is, each neuron, as (*n* is the number of nodes in the network)

$$Z_t = Z_t^{(1)} Z_t^{(2)} \cdots Z_t^{(n)} \ , \quad (25)$$

which allows us to calculate the amount of information processed by each individual node *i* as

$$I^{(i)} = I(Z_t^{(i)} : Z_{t+1}^{(i)}) \ . \quad (26)$$

Note that I omitted the index $t$ on the left-hand-side of Eqs. (24) and (26), assuming that the dynamics of the network becomes stationary as $t \to \infty$, and thus that a sampling of the network states at any subsequent time points becomes representative of the agent's behavior. If the nodes in the network process information independently from each other, then the sum (over all neurons) of the information processed by each neuron would equal the amount of information processed by the entire network. The difference between the two then represents the amount of information that the network processes over and above the information processed by the individual neurons, the *synergistic information*[51]:

$$SI_{\text{atom}} = I(Z_t : Z_{t+1}) - \sum_{i=1}^{n} I^{(i)}(Z_t^{(i)} : Z_{t+1}^{(i)}) \ . \quad (27)$$

---

[b] The notation is slightly modified here to conform to the formalism used in Ref.[51]

The index "atom" on the synergistic information reminds us that the sum is over the indivisible elements of the network: the neurons themselves. As we shall see below, other more general partitions of the network are possible, and often times more appropriate to capture synergy. The synergistic information is related to other measures of synergy that have been introduced independently. One is simply called "integration" and defined in terms of Shannon entropies as[64,66,67]

$$\mathcal{I} = \sum_{i=1}^{n} H(Z_t^{(i)}) - H(Z_t) \ . \quad (28)$$

This measure has been introduced earlier under the name "multi-information"[68,69].

Another measure, called $\Phi_{\text{atom}}$ in Ref.[51], was independently introduced by Ay and Wennekers[70,71] as a measure of the complexity of dynamical systems they called "stochastic interaction", and is defined as

$$\Phi_{\text{atom}} = \sum_{i=1}^{n} H(Z_t^{(i)} | Z_{t+1}^{(i)}) - H(Z_t | Z_{t+1}). \quad (29)$$

Note the similarity between Eqs. (27-29): while (27) measures synergistic information, (28) measures "synergistic entropy" and (29) synergistic conditional entropy in turn. The three are related because entropy and information are related, as for example in Eqs. (11) and (21). Using this relation, it is easy to show that[51]

$$\Phi_{\text{atom}} = SI_{\text{atom}} + \mathcal{I} \ . \quad (30)$$

While we can expect that measures such as (28-30) quantify some aspects of information integration, it is likely that they overestimate the integration because it is possible that elements of the computation are performed by groups of neurons that together behave as a single entity. In that case, subdividing the whole network into independent neurons may lead to the double-counting of integrated information. A cleaner (albeit computationally much more expensive) approach is to find a partition of the network into non-overlapping groups of nodes (parts) that are as independent of each other (information-theoretically speaking) as possible. If we define the partition $P$ of a network into $k$ parts via $P = \{P^{(1)}, P^{(2)}, \cdots P^{(k)}\}$, where each $P^{(i)}$ is a part of the network (an non-empty set of neurons with no overlap between the parts), then we can define a quantity that is analogous to Eq. (29) except that the sum is over the parts rather than the individual neurons[61]

$$\Phi(P) = \sum_{i=1}^{n} H(P_t^{(i)} | P_{t+1}^{(i)}) - H(P_t | P_{t+1}) . \quad (31)$$

In Eq. (31), each part carries a time label because every part takes on different states as time proceeds. The so-called "Minimum Information Partition" (or MIP) is found by minimizing a *normalized* Eq. (31) over all partitions

$$\text{MIP} = \arg \min_{P} \frac{\Phi(P_t)}{N(P_t)} , \quad (32)$$

where the normalization $N(P_t) = (k-1)\min_i[H_{\max}(P_t^{(i)})]$ balances the parts of the partition[62]. Using this MIP, the integrated information $\Phi$ is then simply given by

$$\Phi = \Phi(P = \text{MIP}) . \quad (33)$$

Lastly, we need to introduce one more concept in order to measure information integration in realistic evolving networks. Because the $\Phi$ of a network with a single (or more) disconnected nodes vanishes (because the MIP for such a network is always the partition into the connected nodes in one part, and the disconnected in another) we should attempt to define the computational "main complex", which is that subset of nodes for which $\Phi$ is maximal[62]. This measure will be called $\Phi_{MC}$ in the following.

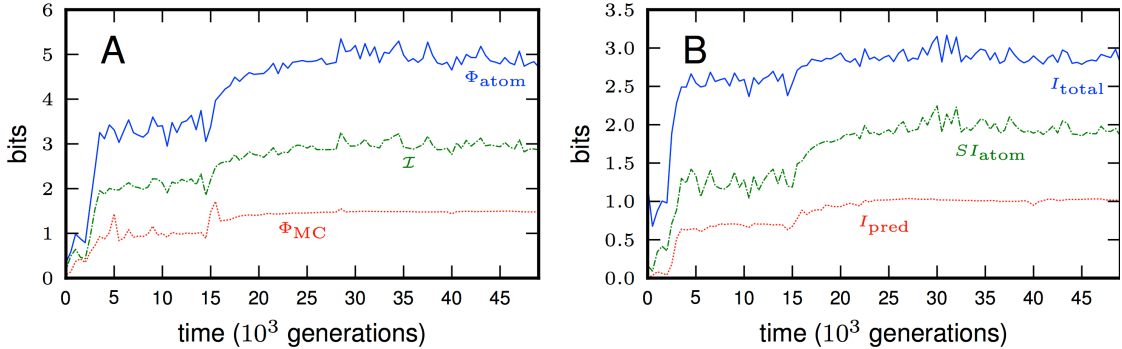

**FIG. 6** A: Three candidate measures of information integration $\Phi_{\text{atom}}$ (29), $\Phi_{MC}$, and $\mathcal{I}$ (28) along the line of descent of a representative evolutionary run in which animats adapted to solve a two-dimensional maze. B: Three measures of information processing, in the same run. Blue: total information $I_{\text{total}}$ (24), green: atomic information $SI_{\text{atom}}$ (27), and red: predictive information $I_{\text{pred}}$ (21) (from Ref.[51]).

While all these measures attempt to capture synergy, it is not clear whether any of them correlate with fitness when an agent evolves, that is, it is not clear whether synergy or integration capture an aspect of the functional complexity of control structures that goes beyond the predictive information defined earlier. To test this, Edlund et al. evolved animats that learned, over 50,000 generations of evolution, to navigate a two-dimensional maze[51], constructed in such a way that optimal navigation requires memory. While measuring fitness, they also recorded six different candidate measures for brain complexity, among which the predictive information Eq. (21), the total information Eq. (24), the synergistic information Eq. (27), as well as the integration Eq. (28), the "atomic

$\Phi''$ (29), as well as the computationally intensive measure $\Phi_{MC}$. Fig. 6 shows a representative run (of 64) that shows the six candidate measures as a function of evolutionary time measured in generations. During this run, the fitness increased steadily, with a big step around generation 15,000 where this particular animat evolved the capacity to use memory for navigation (from Ref.[51]).

It is not clear from a single run which of these measures best correlates with fitness. If we take the fitness attained at the end of each of 64 runs and plot it against the fitness (here measured as the percentage of the achievable fitness in this environment), the sophisticated measure $\Phi_{MC}$ emerges as the clear winner, with a Spearman rank correlation coefficient with achieved fitness of R=0.937 (see Fig. 7). This suggests that measures of information integration can go beyond simple "reactive" measures such as $I_{pred}$ in characterizing complex behavior, in particular when the task requires memory, as was the case there.

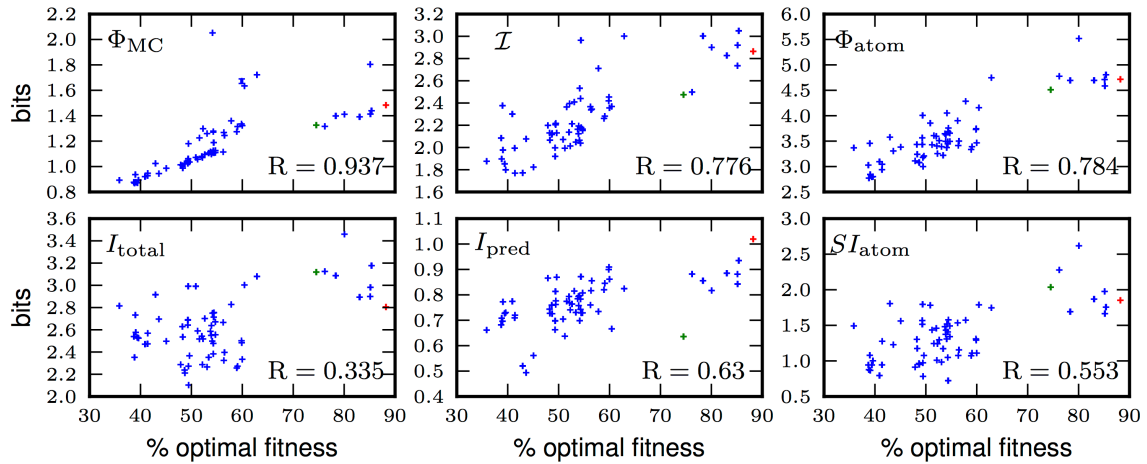

**FIG. 7**: Correlation of information-based measures of complexity with fitness. $\Phi_{MC}$, $\mathcal{I}$, $\Phi_{atom}$, $I_{total}$, $I_{pred}$, and $SI_{atom}$ as a function of fitness at the end of each of 64 independent runs. R indicates Spearman's rank correlation coefficient. The red dot shows the run depicted in Fig. 6 (from Ref.[51]).

## 5. Future Directions

Needless to say, there are many more uses for information theory in evolutionary biology than reviewed here. For example, it is possible to describe the evolution of drug resistance in terms of loss, and subsequent gain, of information: when a pathogen is treated with a drug, the fitness landscape of that pathogen is changed (often dramatically) and as a consequence the genomic sequence that represented information before the administration of the drug is not information (or much less information) about the new environment[22]. As the pathogen adapts to the new environment, it acquires information about that environment and its fitness increases commensurately.

Generally speaking, it appears that there is a fundamental law that links information to fitness (suitably defined). Such a relationship can be written down explicitly for specific systems, such as the relationship between the information content of DNA binding sites with the affinity the binding proteins have with that site[72], or the relationship between the information content of ribozymes and their catalytic activity[73]. We can expect such a relationship to hold as long as information is valuable, and this will always be the case as long as information can be used in decision processes (broadly speaking) that increase the long-term of success of a lineage. It is possible to imagine exceptions to such a law where information would be harmful to an organism, in the sense that signals perceived by a sensory apparatus overwhelm, rather than aid, an organism. Such a situation could arise when the signals are unanticipated, and simply cannot be acted upon in an appropriate manner (for example in animal development). It is conceivable that in such a case mechanisms will evolve that *protect* an organism from signals: this is the basic idea behind the evolution of canalization[74], which is the capacity of an organism to maintain its phenotype in the face of genetic and environmental variation. I would like to point out, however, that strictly speaking, canalization is the evolution of robustness with respect to entropy (noise), not information. If a particular signal cannot be used in order to make predictions, then this signal is not information. In that respect, even the evolution of canalization (if it increases organismal fitness) increases the amount of information an organism has about its environment, because insulating itself from certain forms of noise will increase the reliability of the signals that the organism can use to further its existence.

An interesting example that illustrates the benefit of information and the cost of entropy is the evolution of cooperation, couched in the language of evolutionary game theory[75]. In evolutionary games, cooperation can evolve as long as the decision to cooperate benefits the group more than it costs the individual[76-78]. Groups can increase the benefit accruing to them if they can choose judiciously who to interact with. Thus, acquiring information about the game environment (in this case, the other players) increases the fitness of the group via mutual cooperative behavior. Indeed, it was shown recently that cooperation can evolve among players that interact via the rules of the so-called "Prisoner's Dilemma" game if the strategies that evolve can take into account information about how the opponent is playing[79]. However, if this information is marred by noise (either from genetic mutations that decouple the phenotype from the genotype or from other sources), the population will soon evolve to defect rather than to cooperate. This happens because when the signals cannot be relied upon anymore, information (as the noise increases) gradually turns into entropy. In that case, canalization is the better strategy, and players evolve genes that ignore the opponent's moves[79]. Thus, it appears entirely possible that an information-theoretic formulation of inclusive fitness theory (a theory that predicts the fitness of groups[76,77] that goes beyond Hamilton's kin selection theory) will lead to a predictive framework in which reliable communication is the key to cooperation.

## 6. Conclusions

Information is the central currency for organismal fitness[80], and appears to be that which increases when organisms adapt to their niche[13]. Information about the niche is stored in genes, and used to make predictions about the future states of the environment. Because fitness is higher in well-predicted environments (simply because it is easier to take advantage of the environment's features for reproduction if they are predictable), organisms with more information about their niche are expected to outcompete those with less, suggesting a direct relationship between information content and fitness within a niche (comparisons of information content across niches, on the other hand, are meaningless because the information is not about the same system). A very similar relationship, also enforced by the rules of natural selection, can be found for information acquired not through the evolutionary process, but instead via an organism's sensors. When this information is used for navigation, for example, then a measure called "predictive information" is a good proxy for fitness as long as navigation is performed taking only sensor states into account: indeed, appropriate behavior evolves when information, not fitness, is maximized. If instead decisions are also influenced by memory, different information-theoretic constructions based on the concept of "integrated information" appear to correlate better with fitness, and capture how the brain forms more abstract representations of the world[81] that are used to predict the states of the world on temporal scales much larger than the immediate future. Thus, the ability of making predictions about the world that range far into the future may be the ultimate measure of functional complexity[82] and perhaps even intelligence[83].

**Acknowledgements**

I would like to thank Matthew Rupp for collaboration in work presented in section 3, as well as J. Edlund, A. Hintze, N. Chaumont, G. Tononi, and C. Koch for stimulating discussions and collaboration in the work presented in section 4. This work was supported in part by the Paul G. Allen Family Foundation, The Cambridge Templeton Consortium, the National Science Foundation's Frontiers in Integrative Biological Research Grant FIBR-0527023, as well as NSF's BEACON Center for the Study of Evolution in Action under contract No. DBI-0939454.